# Evolution of photo-excited carrier distribution from anisotropic to isotropic and isotropic photon absorption in graphene

Xiao-Qing Yan[1,2], Zhi-Bo Liu[1,*], Jun Yao[1], Xin Zhao[1], Xu-Dong Chen[1], Yongsheng Chen[2],

Jian-Guo Tian[1,*]

[1]*The Key Laboratory of Weak Light Nonlinear Photonics, Ministry of Education, Teda Applied Physics School, and School of Physics, Nankai University, Tianjin 300457, China*
[2]*The Key Laboratory of Functional Polymer Materials and Center for Nanoscale Science & Technology, Institute of Polymer Chemistry, College of Chemistry, Nankai University, Tianjin 300071, China*

**Abstract:** Femtosecond time-resolved spectroscopy using 400 nm-pump and 800 nm-probe in CVD-grown multilayer graphene provides strong evidence for isotropic distribution of photoexcited carrier after initial relaxation. Indicative of such isotropic distribution is a pump polarization independence of differential reflectivity ($\Delta R/R$) and transmittance ($\Delta T/T$) from pump-probe measurements. Combined with results using 800 nm-pump in [arXiv. 1301.1743v3 (2013)], these pump polarization dependences of time-resolved spectroscopy corroborates the evolution of photo-excited carrier distribution from anisotropic to isotropic with carrier relaxation. And, the absorbance of graphene is identical for in-plane and out-of-plane optical fields. No matter the carrier distribution in momentum space, the influence of carrier on in-plane and out-of-plane optical fields from state filling effect is identical. The sign reversing of ps dynamics signal in graphene/graphite should not directly relate to carrier.

# 1. Introduction

The optical response of graphene, including absorption and photoemission, greatly relates to the photo-excited carrier and carrier relaxation[1,2]. In recent years, the study on optical response and carrier dynamics of graphene has received much attention[3-30]. This 2D material exhibits unique optical properties of interest to the scientist, which make it an important candidate for various electronic and optical applications. Despite substantial progress, the carrier dynamics make physicist fell confused for long time and arise lots of controversy[31], especially for the carrier cooling process.

Based on the anisotropic optical absorption cross sections of carbon nanotube formed by rolling up an graphene sheet[32], the anisotropic optical absorption of graphene could be deduced. However, there has been no experimental study on the anisotropic optical reponse of graphene to verify this deduction. Prior optical measurements have been conducted with normal incident light[10,13], or isotropic optical properties have been regarded for graphene with oblique incident beam[33,34].

---

[*] Corresponding authors: rainingstar@nankai.edu.cn, jjtian@nankai.edu.cn

Malic *et al* stated from microscopic approach based on a many-particle density-matrix framework that carrier distribution is anisotropic in momentum space after generation and the distribution could become isotropic with carrier relaxation. Sun *et al* obtained similar conclusion based on theoretical simulation by solving microscopic kinetic Bloch equation[23]. However, the time-resolved optical pump-probe measurement predicted from the theoretical works contradicts experimental results using normal light as probe beam[11, 26]. In addition, the influence of carrier on optical field of out-of-graphene plane (i.e., c axis) is not known.

In this work, we present a time-resolved ΔR/R and ΔT/T measurements using pump excitation at 400 nm and probe radiation at 800 nm in CVD-grown multilayer graphene, with particular focus on the polarization dependence and the influence of wave vector of probe beam on the experimental signal. Remarkable probe polarization and wave vector dependence of experimental signal indicates the origin of sign reversing of ps dynamics signal should not directly relate to carriers.

The most striking observation-the pump polarization independence of experimental signal-indicates that the optical absorbance of graphene is identical for in-plane and out-of-plane optical fields. This and a comparison with experiments under 800 nm light pumping in Ref. [35] provide solid evidence for carrier distribution from anisotropic after generation to isotropic after enough relaxation, as recently predicted by simulation in Refs. [23, 28, 29]. In the other hand, the experiments demonstrate that the absorbance for in-plane and out-of-plane optical fields are isotropic. To our knowledge, this work presents the first experimental verification of anisotropy of photo-excited carrier distribution and isotropic photon absorption. Because of limitations of theoretical ability of authors, this paper unfortunately cannot provide a full understanding of physical mechanism of the evolution of carrier distribution in graphene.

The paper is organized as follows: In section 2, the experimental arrangement and measurement system are described. In section 3, the experiment results under 400 nm light pumping is presented, including polarization and pump fluence dependence for the cases of TIR and nTIR. The main part of the work is presented in section 4, where we investigate in detail the evolution of photo-excited carrier distribution, signal origin and physical mechanism for the case of TIR. The conclusions is given in section 5.

## 2. Experiment

The sample was that used in [35]. The CVD multilayer graphene (Graphene Supermarket) was transferred to the inclined plane of a BK 7 right-angle prism. The procedure for sample preparation is as in [36]. Experiments were performed in ultra-clean room (T=20 °C, RH-relative humidity=40%). For time-resolved pump-probe measurements, the measurement system was the same as [37] except the pump beam. The 400 nm-light pump beam ($\tau_{FWHM}$=230±30 fs) was obtained from a BBO crystal+BG39 filter combination. For both probe and pump beams, the pump fluence and polarization were varied by a corresponding λ/2-Glan Talyor prism-λ/2 combination. The arrangements are shown in Fig. 1. the absorbance of graphene for normal incident light at 800 nm and 400 nm is 2.3% and 4.14%, respectively[12, 38].

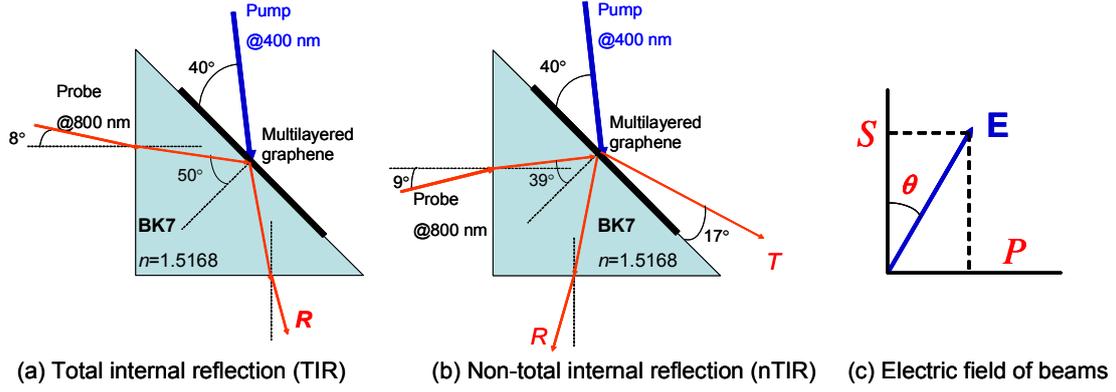

Fig. 1. Experimental setups and definition of polarization of beams.

# 3. Experimental results

## 3.1. TIR

Under s-polarized light pumping, the measured time-resolved differential reflectivity as a function of delay time is shown in Fig. 2a. The pump-induced change of reflectivity around 0 delay time depends on probe polarization and is positive for all polarized probe beam. The peak value of ΔR/R is largest for s-polarized probe beam, and could be decreased to minimum with altering probe polarization to p-polarized (Fig. 2b).

ΔR/R signal decreases immediately after optical excitation. The sign reversal of the pump-probe signal is found for probe beam with polarization of θ=0°, 10° and 30°. And, the delay time of sign reversal increases with probe polarization from S to P. Clearly, the ΔR/R signal of ps dynamics (i.e., the signal with ps decay time, the carrier cooling process) changes greatly with probe polarization. The ΔR/R value at 0.83 delay time increases with probe polarization from negative for S to positive for P (Fig. 2b). Corresponding, the decay time of ps dynamics signal changes greatly with probe polarization (Fig. 2c).

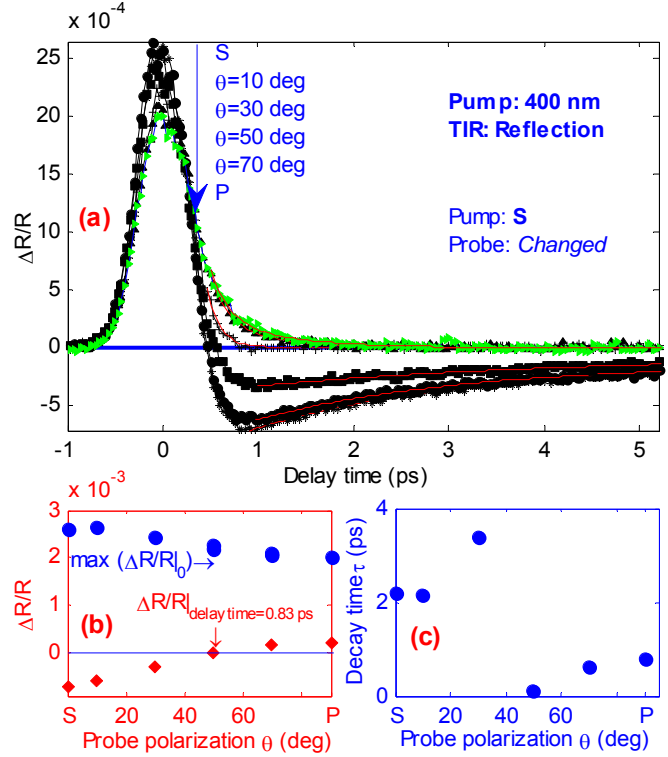

Fig.2. ΔR/R measurement for different polarized probe beam for the case of TIR. (a) The ΔR/R measurement for probe beam with different polarization under pump fluence of 0.045 mJ/cm$^2$. (b) The probe polarization dependences of the peak ΔR/R value and ΔR/R value at 0.83 ps delay time. (c) The decay time of ps dynamics against probe polarization.

Under 400 nm light pumping, the pump-probe signal is pump polarization independent (Fig. 3). The peak ΔR/R value around 0 delay time is linear dependent on pump fluence no matter the probe and pump polarizations. However, when the probe beam is s-polarized, the valley value of ΔR/R decreases with pump fluence at low energy, and then saturated with further increasing pump fluence (Fig. 4).

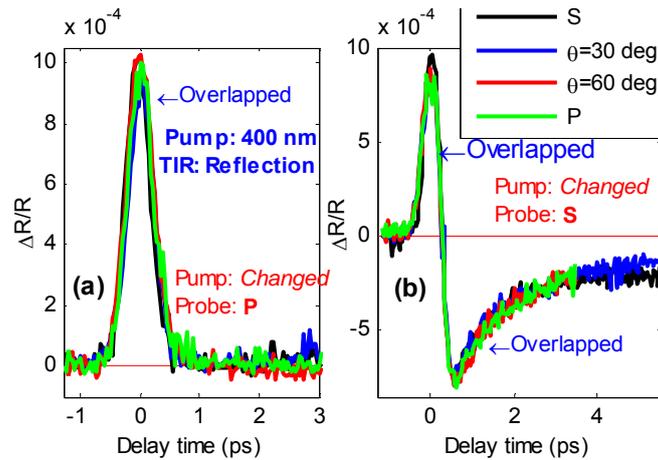

Fig. 3. The ΔR/R measurements under different polarized light pumping beam. (a) probe beam is p-polarized, (b) probe beam is s-polarized.

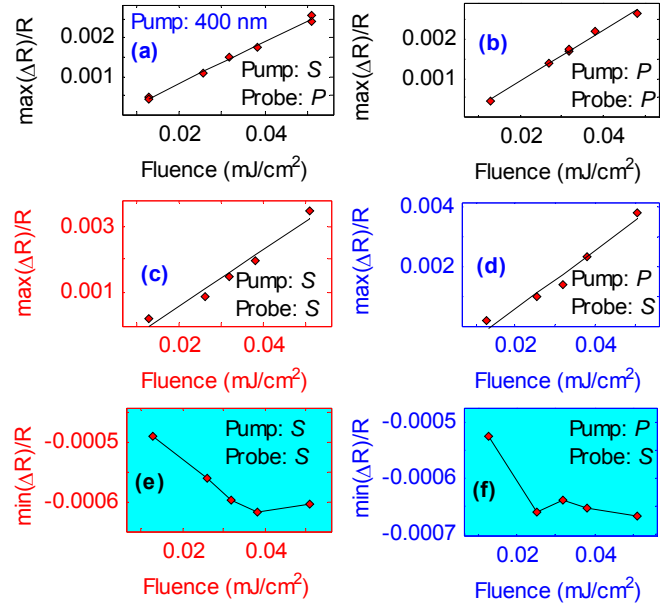

Fig. 4. (a,b,c,d) Linear dependence of the maximum differential reflectivity ΔR/R on pump fluence for different pump-probe combination, the solid line indicates linear fit. (e,f) The dependence of minimum differential reflectivity ΔR/R value around 1 ps delay time on pump fluence.

As shown in Fig. 5, the peak value of ΔR/R is extremely low compared with the valley value of ps dynamics at pump fluence of 7.6 μJ/cm$^2$. With pump fluence increasing, both the peak value and valley value increase. Similar dependence of time scan signal on pump fluence has been observed by Malard *et al* in the transient optical conductivity measurement using pump excitation at 400 nm and probe radiation[12]. In their measurement, both the pump and probe beams were normally focused onto the freely suspended graphene with a 40× objective, the photon-induced transmittance was measured and used to extract transient optical conductivity. Malard *et al* stated from theoretical calculation that the peak signal around 0 delay time originates from inter-band transition and the valley signal at ps dynamics from intra-band transition[12].

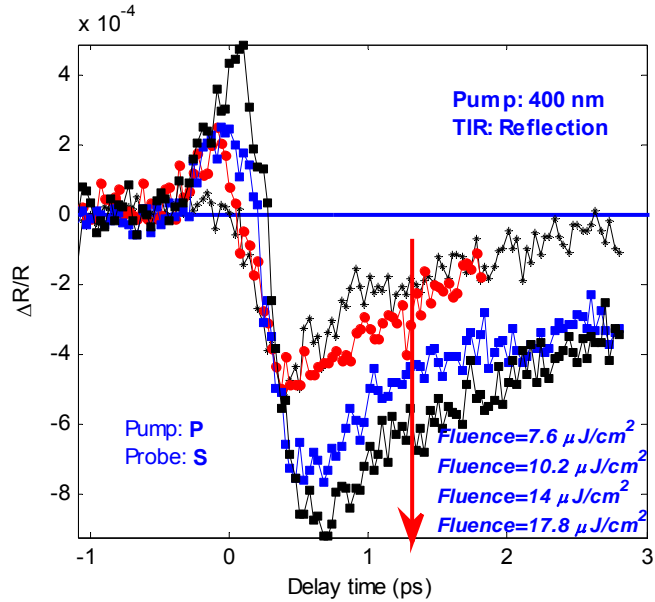

Fig. 5. The differential reflectivity ΔR/R measurement under low pump fluence.

Compared with the experimental results under 800 nm light pumping in [35], the most noticeable difference is the pump polarization independence of ΔR/R.

## 3.2. nTIR: probed with reflected light

Figure 6a shows the time-resolved differential reflectivity ΔR/R. The measured signal changes with probe polarization. The measured ΔR/R is positive from s-polarized light probe and negative for p-polarized probe beam. For probe beam with polarization of θ=68°, the measured ΔR/R signal is negligible. As shown in Fig. 6b, the peak value of ΔR/R could be continuously altered from maximum for s-polarized probe beam to minimum for p-polarized beam. Compared with the case of TIR, no sign reversal in the pump-probe signal is observed. The pump polarization dependence for ΔR/R has not been observed as shown in Fig. 7. The decay time of ps dynamics is probe polarization independent (Fig. 6c). In addition, the ΔR/R signal is linear dependence on pump fluence (Fig. 8) for both s- and p-polarized probe beams.

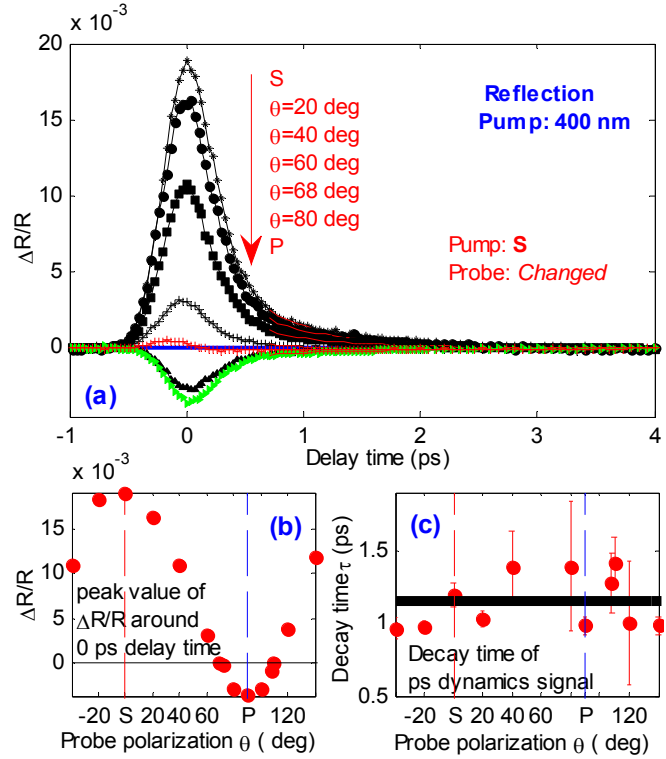

Fig. 6. Probe polarization dependent measurement of ΔR/R for nTIR. (a) The ΔR/R for different polarized probe beam under s-polarized light pumping with fluence of 0.079 mJ/cm$^2$. (b) The probe polarization dependences of the peak/valley (ΔR/R)|$_0$, (c) the probe polarization dependent-decay time of ps dynamics signal.

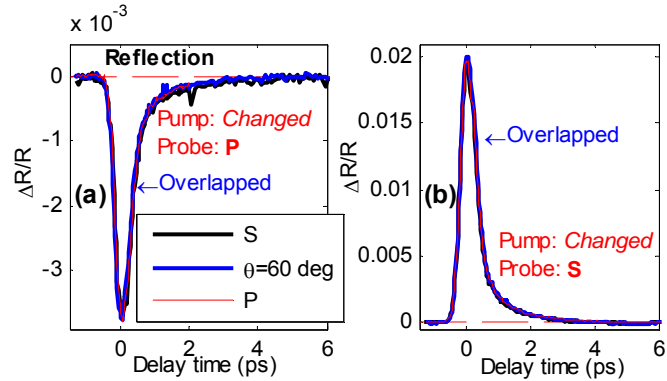

Fig. 7. The differential reflectivity ΔR/R time scan for different polarized pump beam.

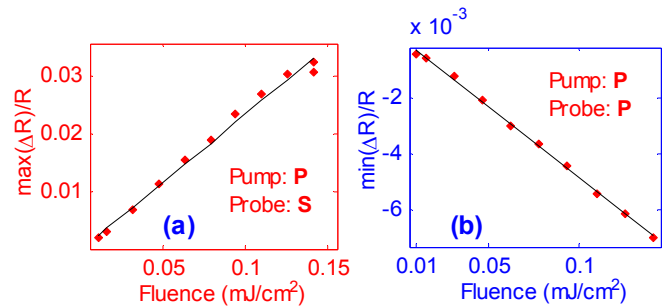
7

Fig. 8. Linear dependence of peak $\Delta R/R|_0$ on pump fluence for different pump-probe polarization combination, solid lines are from linear fits.

## 3.3. nTIR: probed with transmitted light

When the carrier dynamics of graphene was probed with transmitted light, both the measured differential transmittance $\Delta T/T$ and transmittance change $\Delta T$ are probe polarization dependent (Fig. 9). The peak value $\Delta T/T$ and $\Delta T$ around 0 delay time are largest for s-polarized probe beam and smallest for p-polarized probe beam (Figs. 9 and 10). The parameters from fitting are listed in the figure. When the reflected light was used as probe beam, no measured signal was observed for probe beam with $\theta=68°$. However, for probe beam with this polarization, the measured signal is positive.

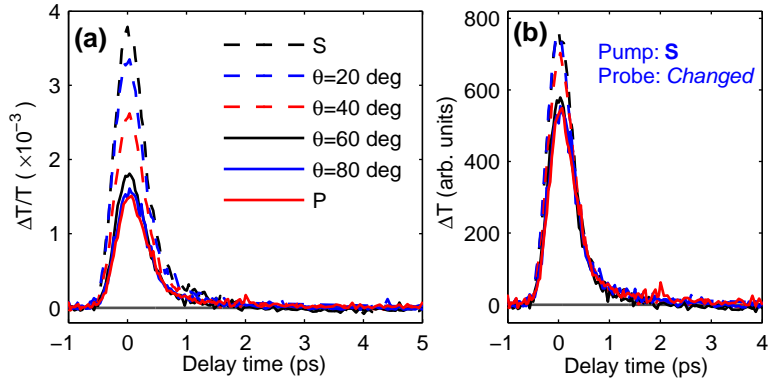

Fig. 9. Polarization dependent $\Delta T/T$ and $\Delta T$ under pump fluence of 0.079 mJ/cm$^2$ for nTIR. The $\Delta T/T$ (a) and transmittance change $\Delta T$ (b) for different polarized probe beam when the pump beam is s-polarized.

As expected, The $\Delta T/T$ signal is pump polarization independent (Figs. 10 and 11). And the peak value of $\Delta T/T$ around 0 delay time depends linearly on pump fluence for both s- and p-polarized pump beams (Fig. 12).

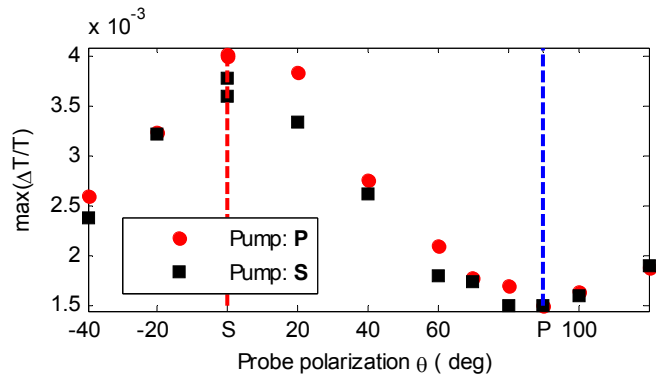

Fig. 10. The peak value of $\Delta T/T$ against probe polarization under s- and p-polarized light pumping.

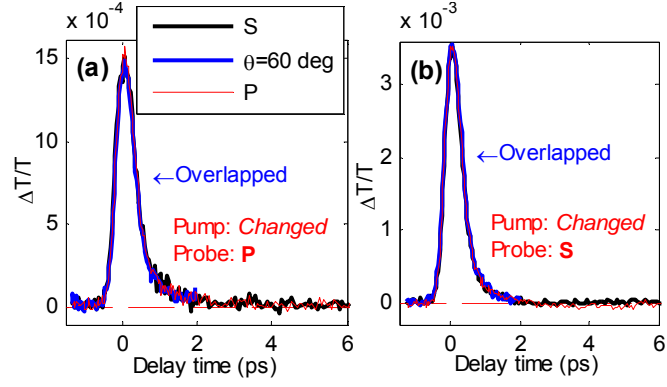

Fig. 11. The differential reflectivity ΔT/T under different polarized light pumping. (a) Probe beam is p-polarized, (b) probe beam is s-polarized.

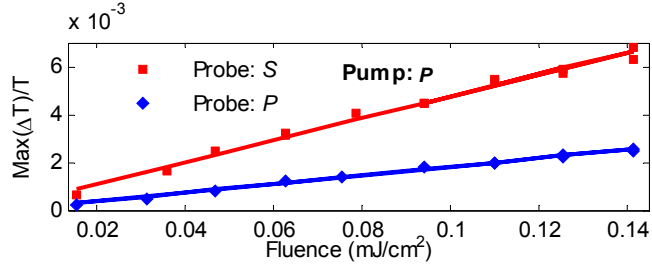

Fig. 12. Linear pump fluence dependence of the peak $\Delta T/T|_0$ for different pump-probe polarization combination, solid lines are from linear fits.

For the three types of probe beam, the signal is probe polarization dependent and pump polarization independent, which is quite different from the case of 800 nm light pumping in [35].

# 4. Discussions

We start our discussion with a brief overview of photo-excited carrier dynamics processes in graphene. During radiation with pump beam, the pump beam excites electrons from valence band to conduction band, as shown in Fig. 13. In energy regime, the generated hot carrier keeps the distribution of the pump pulse intensity (Fig. 18a) and this distribution $n(E)$ is not stable[29]. The photo-excited carrier distribution is transforms from athermal to quasie-quilibrium via carrier-carrier (c-c) scattering and carrier-optical phonon (c-op) scattering processes, and establishe separate Fermi-Dirac quasi-equilibrium distributions with a high temperature $T$ for electrons in the conduction and holes in the valence band[4, 5]. This procedure is generally called carrier thermalization. After thermalization, these hot carriers are still with high energy (i.e., high $T$ in Fermi-Dirac distribution). Following, these hot carriers relax from high energy state to low energy state through carrier-phonon (c-p) scattering (including optical phonon and acoustic phonon), the temperature $T$ used to describe the Fermi-Dirac distribution of hot carrier decreases with carrier relaxation, this procedure is called carrier cooling[29]. After carrier cooling, carriers lose most energy and reach the energy state near Dirac K(K'). At last, electron and hole recombine through Dirac point and get to the equilibrium[5]. Carrier diffusion[12], carrier multiplication[39], population inversion[26] and Auger electron generation[27] may accompany with carrier relaxation.

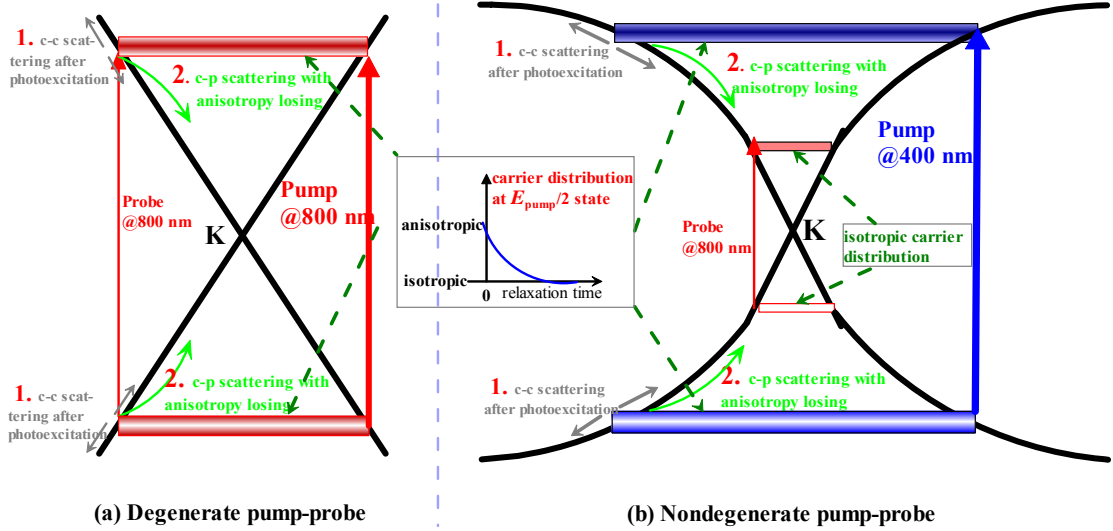

**(a) Degenerate pump-probe**     **(b) Nondegenerate pump-probe**

Fig. 13. Schematics of ultrafast optical interband excitations and carrier relaxation. (a) with pump beam at 800 nm, (b) with pump beam at 400 nm.

As we know, the photon absorption of graphene should meet the requirement of energy and momentum conservations. In graphene, the optical absorption is dominated by intraband transitions at low photon energies (in the far-infrared spectral range) and by interband transitions at higher energies (from mid-infrared to ultraviolet). The intraband and interband transitions in graphene can be modified by doping[40]. The photo-excited carriers could be regarded as transient doping at corresponding energy state, furthermore, affecting the intraband and interband transitions of graphene.

In time-resolved pump-probe measurement, the absorption of probe beam was reduced by the carriers at the energy state of half of probe photon due to Pauling blocking[14]. The carriers are from the excitation of pump beam (Fig. 13). In degenerate pump-probe measurement, the carriers at half of probe photon energy state is without undergoing relaxation; in nondegenerate pump-probe measurement, the carriers at half of probe photon energy state have experienced relaxation. The absorption change of probe beam directly relates to the carrier population at the corresponding energy state. Therefore, information about carrier population could be achieved from the absorption change of probe beam with corresponding photon energy. Therefore, photo-induced transient absorption (pump-probe) experiment is good method to monitor carrier population and its evolution. And, not matter the polarization and wave vector of probe beam the probe beam does not change the carrier dynamics process. In measurements, the electric field components in-graphene plane and out-of-graphene plane are adjusted when the polarization and wave vector of beams are changed. So, the polarization characteristic of graphene ultrafast optical response reflects the anisotropic interaction between optical field and photo-excited carriers.

## 4.1. Evolution of photo-excited carrier distribution

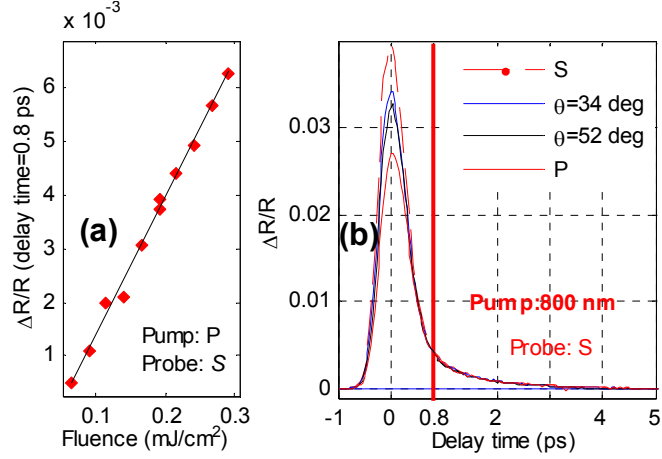

Fig. 14. (a) pump fluence dependence of ΔR/R signal at delay time of 0.8 ps. (b) pump polarization dependence of ΔR/R time scan curves.

First, we analyze the dependence of signal on pump polarization. Owing to the photo-induced signal is from one photon absorption process, the signal is proportional to the carrier population. This dependence has been verified in degenerate pump-probe measurements. In our nondegenerate pump-probe measurement (pump@400 nm here), the probe beam monitors the carriers after thermalization. At this case, the linear dependence of photo-induced signal on pump fluence has been observed as shown in Figs. 4a-d, 8 and 12. It means that the carrier multiplication at so high energy state has not taken place. In other words, the carrier amount does not change during initial relaxation. And, the contribution of intra-band absorption to absorption change of probe beam could be eliminated.

The pump polarization independence of signal reflects that the carrier population at probe photon energy state should be the same for each polarized pump beam. Furthermore, the photo-excited carrier amount is identical for each polarized pump beam due to carrier population conservation during initial relaxation. So, the photon absorption coefficient of graphene at 400 nm is identical for ab- and c-optical field components.

In [35], we find that the pump-probe signal greatly depends on pump polarization in degenerate pump-probe measurement (Figs. 2b, 2d, 3d and 3e). However, the signal with pump polarization dependence is the signal around 0 delay time, i.e., the periods of carrier generation and thermalization. Contrary, the signal of ps dynamics is pump polarization independent (Figs. 14-16).

In addition, the signal of ps dynamics, which reflects the carrier population after thermalization at state of half of pump photon energy in degenerate pump-probe measurement, is nearly linear dependent on pump fluence. As we know, the pump fluence could linearly determines the amount of photo-excited carrier at relatively low pump fluence no matter the polarization of pump beam. Thus, the amount of carrier generated by different polarized pump beam should be identical because ps dynamics signal is pump polarization independent. So, the photon absorption coefficient of graphene at 800 nm is identical for ab- and c-optical field components too.

Based on the broadband optical absorption coefficient of graphene and band structure of graphene[38, 40], the conclusion of identical absorption for ab- and c-optical field components in graphene could be predicted.

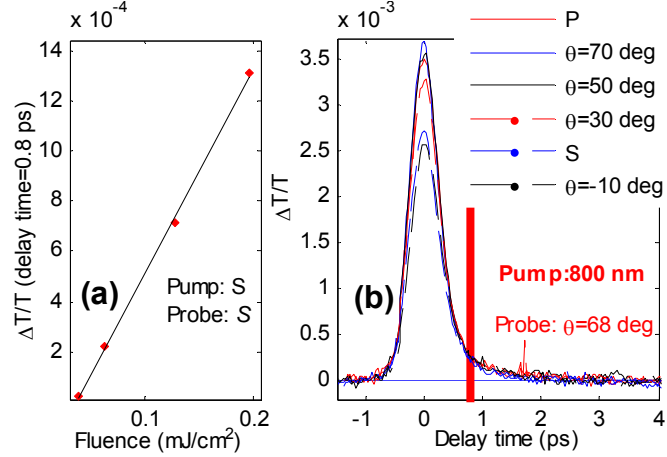

Fig. 15. (a) pump fluence dependence of ΔT/T signal at delay time of 0.8 ps. (b) pump polarization denpendence of ΔT/T time scan curves.

Since amount of carrier induced by different polarized pump beam is the same, the pump polarization dependence of signal around 0 delay time should not originate from anisotropy of photon absorption. It should be from anisotropic distribution of carrier for different polarized pump beam. Thus, we can conclude that the distribution of photo-excited carrier is anisotropic and relates to the polarization of excitation light.

Based on above analysis, the carrier distribution after an optical excitation and the pump polarization characteristics of pump-probe signal could be understood as:

1) Different polarized pump beam generates identical amount of photo-excited carrier, but the carrier distribution is anisotropic for different polarized pump beam. So, the ΔR/R and ΔT/T at 0 delay time is pump polarization dependent in degenerate pump-probe measurement.

2) Immediately after optical excitation, the carriers relax via c-c and c-p scattering from the state of half of pump photon energy with re-distributing carrier, as shown in Figs. 13 and 18. During this thermalization process, the anisotropy of carrier distribution loses with carrier relaxation. The carriers at the energy state of half of pump photon not only become isotropic but also decrease with relaxation. So, the pump polarization dependence of signal around 0 delay time decreases with carrier relaxation, as shown in Figs. 14b, 15b and 16b.

3) In the carrier cooling process, the carrier distribution has been completely isotropic at every energy state. Accordingly, the ps dynamics signal is without pump polarization dependence (Figs. 14b, 15b and 16b). For the case of 400 nm light pumping, the probe beam monitors the carrier relaxed from high energy state. When the carriers reach the state of half of probe photon energy, the carrier distribution has been isotropic. So, the pump polarization characteristic has nerve been observed in our non-degenerate pump-probe measurements.

According to Malic *et al*'s work, c-p scattering is responsible for the anisotropy vanishing during carrier relaxation. And, about 50 fs is needed for the relaxation of initial non-equilibrium carrier distribution from highly anisotropic to completely isotropic[28].

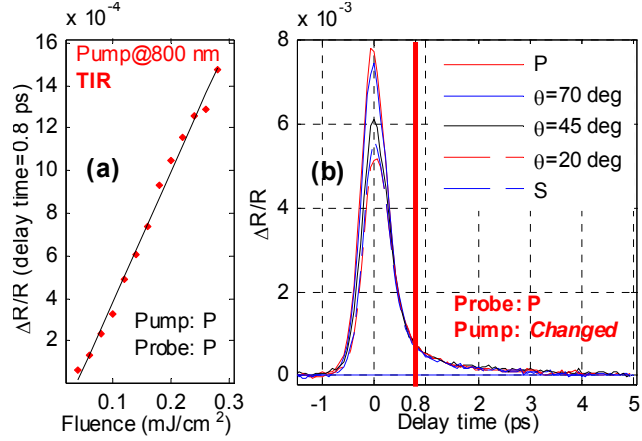

Fig. 16. (a) pump fluence dependence of ΔR/R signal at delay time of 0.8 ps for the case of TIR, pump beam is at 800 nm. (b) pump polarization denpendence of ΔR/R time scan curves for p-polarized probe beam.

## 4.2. Signal probed with total internal reflected light

Let us turn to the case of TIR under 800 nm light pumping , as expected, the signal around 0 delay time depends greatly on pump polarization, and the pump polarization independence has been also observed for ps dynamics signal (Fig. 2d in [35] and Fig. 15 here). The evolution of carrier distribution from anisotropic to isotropic is confirmed again.

Based on above discussion and experiments, it is well known that:
(1) for the case of 400 nm light pumping, the whole signal through the scan time is from isotropic distributed carrier. As shown above, under 400 nm light pumping, the peak $\Delta R/R|_0$ is linearly dependent on pump fluence. And, when the probe beam is s-polarized, the pump fluence dependence of the valley value of ps dynamics signal is identical to that under 800 nm light pumping. At low pump fluence, the valley signal enhances with pump fluence. At high pump fluence, the valley signal is nearly saturated.

(2) for the case of 800 nm light pumping, the signal around 0 delay time is from carrier with anisotropic distribution, the ps dynamics signal is from carrier with isotropic distribution. The linear dependence of peak value of signal around 0 delay time is maintained no matter the polarizations of pump and probe beam (Fig. S3 therein), but it is not available for ps dynamics signal. The ps dynamics signal probed with s-polarized reflected light is not linearly dependent on pump fluence (Figs. S3e and S3f therein). But, the ps dynamics signal linearly depends on pump fluence when the probe beam is p-polarized (Fig. 16a).
Since the carrier distributions for the signal of around 0 delay time pumped with 400 nm light and ps dynamcis signal pumped with 800 nm light are isotropic, the different pump fluence dependence of the signal from s-polarized probe beam indicates that the carrier distribution cloud not be the origin of the valley signal.
Owing to the ps dynamics signal from p-polarized total internal reflected beam is positive, the physical mechanism of the negative ΔR/R signal under total internal reflected beam should be different from that of negative ΔR/R signal from p-polarized non-total internal reflected beam.

Ruzicka *et al* have studied the carrier dynamics of single-layer CVD graphene using spatially and temporally resolved pump-probe spectroscopy. The experiment shows that the sign reversal of the ps dynamics signal relates to the measured position with respected to pumping area. As we know, the Goos–Hänchen shift[41-43] is considerable for total internal reflected light (in the order of wavelength). But, the Goos–Hänchen shift is much smaller than beam waist of probe beam in our measurements. The observed probe polarization dependence of sign reversal in ps dynamics signal could not originate from Goos–Hänchen shift. The reasons are as follows:

(1) the Goos–Hänchen shift difference between s- and p-polarized beam is much smaller than the wavelength for the total internal reflected probe beam in the measurements (<0.1 $\lambda_{probe}$), and is much smaller than the length which is needed (>1 μm) for observing ps dynamics signal from sign reversing to sign keeping (Fig. 2 in Ref. [13]).

(2) If the sign difference of ps dynamics signal between s- and p-polarized probe beam is really caused by Goos–Hänchen shift difference between s- and p-polarized beam. The changing of ps dynamics signal from s- to p-polarized should be similar to the changing of signal around 0 delay time. As shown in Ref. [35], when the pump beam is s-polarized, the signal around 0 delay time increases with probe polarization changing from P to S, and the signal keeps positive. The ps dynamics signal is positive for p-polarized probe beam. If the probe polarization dependence of signal is really caused by Goos–Hänchen shift, the ps dynamics signal should increases with probe polarization from P to S and keeps positive, which is opposite to the experimental result.

(3) The measurement has been performed under different degree of overlapping between probe and pump beams, the ps dynamics signal is always positive for p-polarized probe beam and negative for s-polarized probe beam. The movement of pump beam with respected to probe beam could not change the sign of ps dynamics signal, as shown in Fig. 16.

So, the observed probe polarization dependence of ps dynamics signal is not similar to that reported by Ruzicka *et al* and not caused by Goos–Hänchen shift. It is should be directly related to intrinsic optical response of graphene to total internal reflected light.

As we know, the valley signal from s-polarized probe beam is not proportional to the pump fluence under the two types of pump beam (800 and 400 nm). However, the peak value around 0 delay time linearly depends on pump fluence. It means that another sign origin besides Pauling blocking in ps dynamics signal should not directly relate to carrier.

## 4.3. isotropic ultrafast optical response

Now, we turn to the probe polarization dependence of pump-probe signal. As shown in Figs. 2, 6 and 9, the differential transmittance ΔT/T and reflectivity ΔR/R signal greatly depends on probe polarization in the whole scan time. The probe polarization dependence of signal could be well explained with Maxwell's equation [44], which means the influence of photo-excited carrier on optical field of probe beam is identical. i.e., the ultrafast optical response is isotropic in graphene.

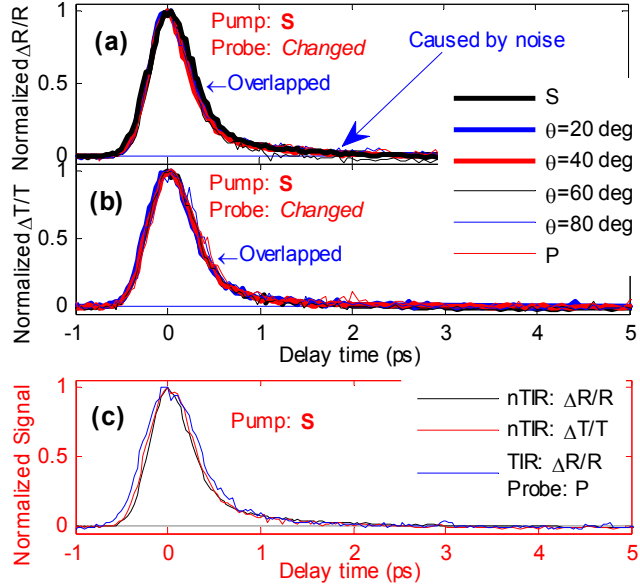

Fig. 17. Normalized differential signal for different polarized probe beam. Pump beam is s-polarized light at wavelength of 400 nm. (a) the normalized ΔR/R from Fig. 6, (b) normalized ΔT/T from from Fig. 9, (c) comparison of normalized signal from different probing ways. Figures (a) and (b) share the same legend.

For the case of nTIR, as shown in Figs. 17-19, the normalized differential signal is identical no matter the probe polarization, and probe beams (reflected light or transmitted light). Furthermore, the delay time of ps dynamics signal is identical. It is reasonable, since the carrier dynamics of graphene is not modified by probe polarization.

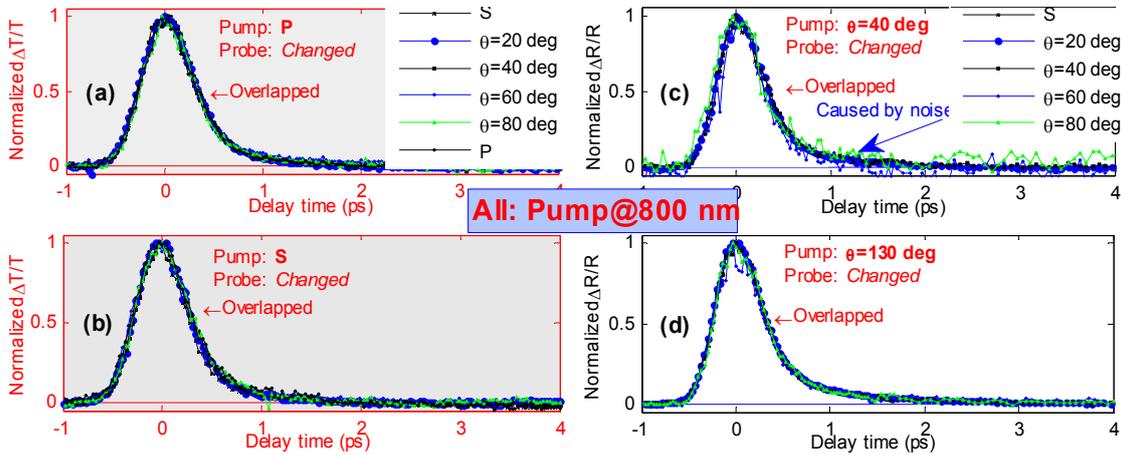

Fig. 18. Normalized differential signal for different polarized probe beam for the case of nTIR. Pump beam is at wavelength of 800 nm. (a) and (b) are probed with transmitted light with p- and s-polarized pump beam, respectively, the two figures share the same legend.
. (c) and (d) are probed with reflected light, the two figures share the same legend.

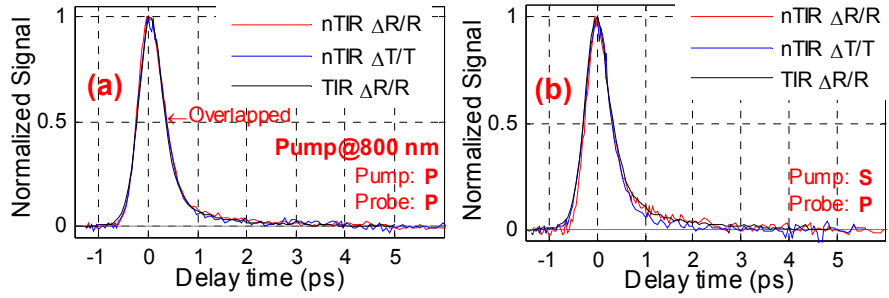

Fig. 19. comparison of normalized signal with different probing way under 800 nm light pumping. (a) pump beam is p-polarized, (b) pump beam is s-polarized. The two figures share the same legend.

# 5. Conclusions

Our experiment has provided the first direct evidence of evolution of carrier distribution from anisotropic to isotropic. The absorption coefficient for optical field in-graphene plane and out-of-graphene plane is identical. Although the carrier initial distribution of photoexcited carrier is anisotropic, the influences of photo-excited carrier on optical fields in-graphene plane and out-of-graphene plane are identical. The sign reversing in ps dynamics signal should not directly relate to carrier, may be related by the structure change of graphene sheet. In addition, the measurement with sign reversing in ps dynamics signal does not reflects the intrinsic carrier dynamics of graphene.

# ACKNOWLEDGMENTS

This work was supported by the Chinese National Key Basic Research Special Fund (grant 2011CB922003), the National Natural Science Foundation of China (grants 11174159, 11374164, 11304166).

*Note added in proof.*—A similar study of the carrier distribution in graphene has recently been posted by Mittendorff *et al*. [45].

# References

1. F. Xia, T. Mueller, Y. M. Lin, A. Valdes-Garcia and P. Avouris, Nat Nanotechnol **4** (12), 839-843 (2009).
2. C. H. Lui, K. F. Mak, J. Shan and T. F. Heinz, Phys Rev Lett **105** (12), 127404 (2010).
3. J. C. Johannsen, S. Ulstrup, F. Cilento, A. Crepaldi, M. Zacchigna, C. Cacho, I. C. E. Turcu, E. Springate, F. Fromm, C. Raidel, T. Seyller, F. Parmigiani, M. Grioni and P. Hofmann, Phys Rev Lett **111** (2), 027403 (2013).

4.  I. Gierz, J. C. Petersen, M. Mitrano, C. Cacho, I. C. E. Turcu, E. Springate, A. Stöhr, A. Köhler, U. Starke and A. Cavalleri, Nat Mater **10.1038/nmat3757** (2013).
5.  T. Limmer, J. Feldmann and E. Da Como, Phys Rev Lett **110** (21), 217406 (2013).
6.  K. Seibert, G. C. Cho, W. Kutt, H. Kurz, D. H. Reitze, J. I. Dadap, H. Ahn, M. C. Downer and A. M. Malvezzi, Physical review. B, Condensed matter **42** (5), 2842-2851 (1990).
7.  M. Breusing, S. Kuehn, T. Winzer, E. Malic, F. Milde, N. Severin, J. P. Rabe, C. Ropers, A. Knorr and T. Elsaesser, Phys Rev B **83** (15), 153410 (2011).
8.  M. Breusing, C. Ropers and T. Elsaesser, Phys Rev Lett **102** (8), 086809 (2009).
9.  D. Sun, C. Divin, C. Berger, W. A. de Heer, P. N. First and T. B. Norris, Phys Rev Lett **104** (13), 136802 (2010).
10. D. Sun, Z. K. Wu, C. Divin, X. B. Li, C. Berger, W. A. de Heer, P. N. First and T. B. Norris, Phys Rev Lett **101** (15), 157402 (2008).
11. D. Sun, The University of Michigan, 2009.
12. M. M. Leandro, M. Kin Fai, A. H. C. Neto, N. M. R. Peres and F. H. Tony, New J Phys **15** (1), 015009 (2013).
13. B. A. Ruzicka, S. Wang, J. W. Liu, K. P. Loh, J. Z. Wu and H. Zhao, Opt Mater Express **2** (6), 708-716 (2012).
14. J. Z. Shang, Z. Q. Luo, C. X. Cong, J. Y. Lin, T. Yu and G. G. Gurzadyan, Appl Phys Lett **97** (16), 163103 (2010).
15. F. Carbone, G. Aubock, A. Cannizzo, F. Van Mourik, R. R. Nair, A. K. Geim, K. S. Novoselov and M. Chergui, Chem Phys Lett **504** (1-3), 37-40 (2011).
16. J. M. Dawlaty, S. Shivaraman, M. Chandrashekhar, F. Rana and M. G. Spencer, Appl Phys Lett **92** (4), 042116 (2008).
17. H. N. Wang, J. H. Strait, P. A. George, S. Shivaraman, V. B. Shields, M. Chandrashekhar, J. Hwang, F. Rana, M. G. Spencer, C. S. Ruiz-Vargas and J. Park, Appl Phys Lett **96** (8), 081917 (2010).
18. S. Butscher, F. Milde, M. Hirtschulz, E. Malic and A. Knorr, Appl Phys Lett **91** (20), 203103 (2007).
19. F. Rana, P. A. George, J. H. Strait, J. Dawlaty, S. Shivaraman, M. Chandrashekhar and M. G. Spencer, Phys Rev B **79** (11), 115447 (2009).
20. R. W. Newson, J. Dean, B. Schmidt and H. M. van Driel, Opt Express **17** (4), 2326-2333 (2009).
21. L. B. Huang, G. V. Hartland, L. Q. Chu, Luxmi, R. M. Feenstra, C. X. Lian, K. Tahy and H. L. Xing, Nano Lett **10** (4), 1308-1313 (2010).
22. S. Tani, F. Blanchard and K. Tanaka, Phys Rev Lett **109** (16), 166603 (2012).
23. B. Y. Sun and M. W. Wu, New J Phys **15** (8), 083038 (2013).
24. B. Y. Sun, Y. Zhou and M. W. Wu, Phys Rev B **85** (12), 125413 (2012).
25. A. Tomadin, D. Brida, G. Cerullo, A. C. Ferrari and M. Polini, Phys Rev B **88** (3), 035430 (2013).
26. T. Li, L. Luo, M. Hupalo, J. Zhang, M. C. Tringides, J. Schmalian and J. Wang, Phys Rev Lett **108** (16), 167401 (2012).
27. K. J. Tielrooij, J. C. W. Song, S. A. Jensen, A. Centeno, A. Pesquera, A. Zurutuza Elorza, M. Bonn, L. S. Levitov and F. H. L. Koppens, Nature Phys. **9** (4), 248-252 (2013).
28. E. Malic, T. Winzer and A. Knorr, Appl Phys Lett **101** (21), 213110 (2012).

29. E. Malic, T. Winzer, E. Bobkin and A. Knorr, Phys Rev B **84** (20), 205406 (2011).
30. T. Winzer, E. Malić and A. Knorr, Phys Rev B **87** (16), 165413 (2013).
31. T. Winzer and E. Malic, Phys Status Solidi B **248** (11), 2615-2618 (2011).
32. Y. Murakami, E. Einarsson, T. Edamura and S. Maruyama, Phys Rev Lett **94** (8), 087402 (2005).
33. G. Pirruccio, L. Martín Moreno, G. Lozano and J. Gómez Rivas, Acs Nano **7** (6), 4810-4817 (2013).
34. Q. Ye, J. Wang, Z. Liu, Z.-C. Deng, X.-T. Kong, F. Xing, X.-D. Chen, W.-Y. Zhou, C.-P. Zhang and J.-G. Tian, Appl Phys Lett **102** (2), 021912 (2013).
35. X.-Q. Yan, Z.-B. Liu, J. Yao, X. Zhao, X.-D. Chen, F. Xing, Y. Chen and J.-G. Tian, arXiv preprint arXiv:1301.1743 (2013).
36. X.-D. Chen, Z.-B. Liu, C.-Y. Zheng, F. Xing, X.-Q. Yan, Y. Chen and J.-G. Tian, Carbon **56** (0), 271-278 (2013).
37. X. Zhao, Z.-B. Liu, W.-B. Yan, Y. Wu, X.-L. Zhang, Y. Chen and J.-G. Tian, Appl Phys Lett **98** (12), 121905 (2011).
38. K. F. Mak, M. Y. Sfeir, Y. Wu, C. H. Lui, J. A. Misewich and T. F. Heinz, Phys Rev Lett **101** (19), 196405 (2008).
39. T. Winzer, A. Knorr and E. Malic, Nano Lett **10** (12), 4839-4843 (2010).
40. K. F. Mak, L. Ju, F. Wang and T. F. Heinz, Solid State Commun **152** (15), 1341-1349 (2012).
41. http://en.wikipedia.org/wiki/Goos%E2%80%93H%C3%A4nchen_effect.
42. Y. Song, H.-C. Wu and Y. Guo, Appl Phys Lett **100** (25), - (2012).
43. Z. H. Wu, F. Zhai, F. M. Peeters, H. Q. Xu and K. Chang, Phys Rev Lett **106** (17), 176802 (2011).
44 Xiao-Qing Yan, et al, (in preparision)
45. Martin Mittendorff, Torben Winzer, Ermin Malic, Andreas Knorr, Claire Berger, Walter A. de Heer, Harald Schneider, Manfred Helm, and Stephan Winnerl, "Anisotropy of excitation and relaxation of photogenerated Dirac electrons in graphene," http://arxiv.org/abs/1312.5860